\documentclass[twocolumn,%
                showpacs,%
                superscriptaddress,%
                longbibliography,%
                amsmath,%
                amssymb,%
                preprint,%
                10pt,%
                showkeys,%
                aps,%
                floatfix]{revtex4-1}
\usepackage[utf8]{inputenc}
\usepackage{graphicx}
\usepackage{dcolumn}
\usepackage{bm}
\usepackage{hyperref}
\usepackage{amsthm, amsmath, mathtools}
\usepackage{physics}
\usepackage{wasysym}

\usepackage{kantlipsum}
\usepackage{xspace}
\usepackage{siunitx}
\usepackage{multirow}
\usepackage[normalem]{ulem}
\usepackage[usenames,dvipsnames]{color}
\usepackage{cleveref}
\usepackage{enumitem} 

\makeatletter
\newcommand*{\addSI}{%
  \close@column@grid
  \cleardoublepage
  \twocolumngrid
}
\makeatother

\def\eqref#1{(\ref{#1})}

\newcommand{\un}[1]{\,\mathrm{#1}}

\newcommand{\kb}{k_{\text{\tiny B}}}

\newcommand{\ie}{\emph{i.e.}\xspace}
\newcommand{\eg}{\emph{e.g.}\xspace}

\newcommand{\Z}{\mathbb{Z}}

\newcommand{\lb}{lp}

\newcommand{\qe}{\textsc{Quantum~ESPRESSO}\xspace}
\newcommand{\wannier}{\textsc{Wannier90}\xspace}

\newtheorem*{theorem*}{Theorem}
\newtheorem*{corollary*}{Corollary}

\usepackage[normalem]{ulem}
\usepackage[usenames,dvipsnames]{color}

\definecolor{verde}{rgb}{0.,0.6,0}
\definecolor{rosso}{rgb}{0.9,0.0,0.2}
\definecolor{magenta}{rgb}{0.9,0.2,0.9}

\newcommand{\editor}[2]{%
  \expandafter\newcommand\csname #1note\endcsname[1]{%
    \textcolor{#2}{(\textbf{#1:} ##1)}}%
  \expandafter\newcommand\csname #1\endcsname[1]{%
    \textcolor{#2}{##1}}%
  \expandafter\newcommand\csname #1cancel\endcsname[1]{%
    \textcolor{#2}{\sout{##1}}}%
  \expandafter\newcommand\csname #1change\endcsname[2]{%
    \textcolor{#2}{\sout{##1} ##2}}%
  \newenvironment{#1text}{\color{#2}}{\color{black}}
}

\definecolor{tangerine}{rgb}{0.944,0.522,0}
\editor{SB}{tangerine}
\editor{PP}{verde}
\editor{FG}{rosso}
\editor{resub}{cyan}
\editor{highlight}{magenta}

\begin{document}

\title{Oxidation states, Thouless' pumps, and non-trivial ionic \\ transport in non-stoichiometric electrolytes}

\author{Paolo Pegolo}
\affiliation{SISSA -- Scuola Internazionale Superiore di Studi Avanzati, 34136 Trieste, Italy}
\author{Federico Grasselli}
\altaffiliation{Current affiliation: COSMO -- Laboratory of Computational Science and Modelling, IMX, \'Ecole Polytechnique F\'ed\'erale de Lausanne, 1015 Lausanne, Switzerland}
\affiliation{SISSA -- Scuola Internazionale Superiore di Studi Avanzati, Trieste, Italy}
\author{Stefano Baroni}\email{baroni@sissa.it}
\affiliation{SISSA -- Scuola Internazionale Superiore di Studi Avanzati, Trieste, Italy}
\affiliation{CNR -- Istituto Officina dei Materiali, SISSA, 34136 Trieste}

\date{\today}

\begin{abstract}
    Thouless' quantization of adiabatic particle transport permits to associate an integer topological charge with each atom of an electronically gapped material. If these charges are additive and independent of atomic positions, they provide a rigorous definition of atomic oxidation states and atoms can be identified as integer-charge carriers in ionic conductors. Whenever these conditions are met, charge transport is necessarily convective, \ie~it cannot occur without substantial ionic flow, a transport regime that we dub \emph{trivial}. We show that the topological requirements that allow these conditions to be broken are the same that would determine a Thouless' pump mechanism if the system were subject to a suitably defined time-periodic Hamiltonian. The occurrence of these requirements determines a \emph{non-trivial} transport regime whereby charge can flow without any ionic convection, even in electronic insulators.
    These results are first demonstrated with a couple of simple molecular models that display a quantum pump mechanism upon introduction of a fictitious time dependence of the atomic positions along a closed loop in configuration space. We finally examine the impact of our findings on the transport properties of non-stoichiometric alkali-halide melts, where the same topological conditions that would induce a quantum pump mechanism along certain closed loops in configuration space also determine a non-trivial transport regime such that most of the total charge current results to be uncorrelated from the ionic ones.
\end{abstract}

\pacs{      
  66.10.-x	
  61.20.Ja	
}

\keywords{ 
  Oxidation states, Topological quantum numbers, Thouless' charge pump, Adiabatic dynamics%
}

\maketitle

\section{Introduction}
Atomic oxidation states (OS) are ubiquitous in chemistry and widely used to describe redox reactions, electrolysis, and many electro-chemical processes. In spite of their fundamental nature, OSs have long eluded a proper quantum-mechanical interpretation. As a matter of fact, the commonly accepted definition provided by IUPAC (\emph{OS of an atom is the charge of this atom after ionic approximation of its heteronuclear bonds}~\cite{IUPAC}) can hardly be given a rigorous quantitative meaning. As one sees, this statement stands on \emph{approximating} a \emph{real} number that expresses a \emph{static} property (the atomic charge, which is not even well-defined quantum-mechanically) to the closest integer, a procedure that is intrinsically ill-defined and potentially misleading in some cases~\cite{Raebiger2008,resta2008}. This predicament has been reversed by a recent paper~\cite{grasselli2019} where, building on previous work~\cite{rappe2012} based on the modern theory of polarization~\cite{Resta2007,vanderbilt2018berry}, it was shown that OSs can indeed be defined as topological quantum numbers~\cite{thouless83} describing the charge \emph{dynamically} displaced by individual atoms along closed paths in atomic configuration space, under periodic boundary conditions. While such a dynamical definition may not be expedient to describe charge-ordering effects~\cite{Pickett2014}, it is indeed expected to fit transport theory, where the dipole displaced along an atomic trajectory is actually all is needed to define and compute the electrical conductivity. In fact, by leveraging a recently established \emph{gauge invariance} principle of transport coefficients~\cite{aris-nature,ercole2016gauge}, it was also shown that, whenever topological charges are both additive and independent of atomic positions, atoms can be identified as integer carriers in adiabatic charge transport and the latter is purely convective, \ie it can only occur along with the displacement of the atomic charge carriers. We call this transport regime \emph{trivial}. Of course, chemically relevant situations occur where different atoms of a same species feature different OSs, depending on the local chemical environment. Although the occurrence of such cases is indeed compatible with a comprehensive topological definition of OSs~\cite{rappe2012}, breaking the additivity and position-independence of atomic OSs has important consequences on transport properties, the most perspicuous of which is the possibility that charge may be displaced without any associated atomic convection, a ionic transport regime that we dub \emph{non-trivial}.

In this work we address the topological conditions to be met in order to break the additivity and position independence of topological charges. We show that, when these conditions occur, it is possible to identify closed paths in atomic configuration space such that, if the system is adiabatically driven along one of these paths, a quantum-pump mechanism determines the net displacement of an integer charge, not corresponding to any  ionic displacements. These findings are illustrated with a couple of toy molecular models dealt with in periodic boundary conditions (PBC): the positive hydrogen trimer ion, $\mathrm H^+_3$, which displays a lack of topological-charge additivity, and the $\mathrm{K_3Cl}$ neutral complex, where different K atoms feature different and non additive topological charges.
We finally consider a non-stoichiometric $\mathrm {K}_{x}(\mathrm{KCl})_{1-x}$ ionic melt, which, for small enough $x$, features a finite electron-energy gap. In this case, the same topological conditions responsible for the quantum-pump mechanisms in the molecular models considered above determine a non-trivial transport regime, whereby the total charge current mainly results from lone pairs of electrons donated by the excess metal atoms and whose motion is largely uncorrelated from the ionic ones.
Non-stoichiometric molten salts are paradigmatic cases of systems featuring solvated electrons, other notable examples including non-stoichiometric electrolytes and metal solutions in a ionic solvent, to which we believe that the bulk of our analysis also applies~\cite{larsen2010, symons1976solutions,shkrob2006}.

\section{theory}
\begin{figure}
    \centering
    \includegraphics[width=0.9\columnwidth]{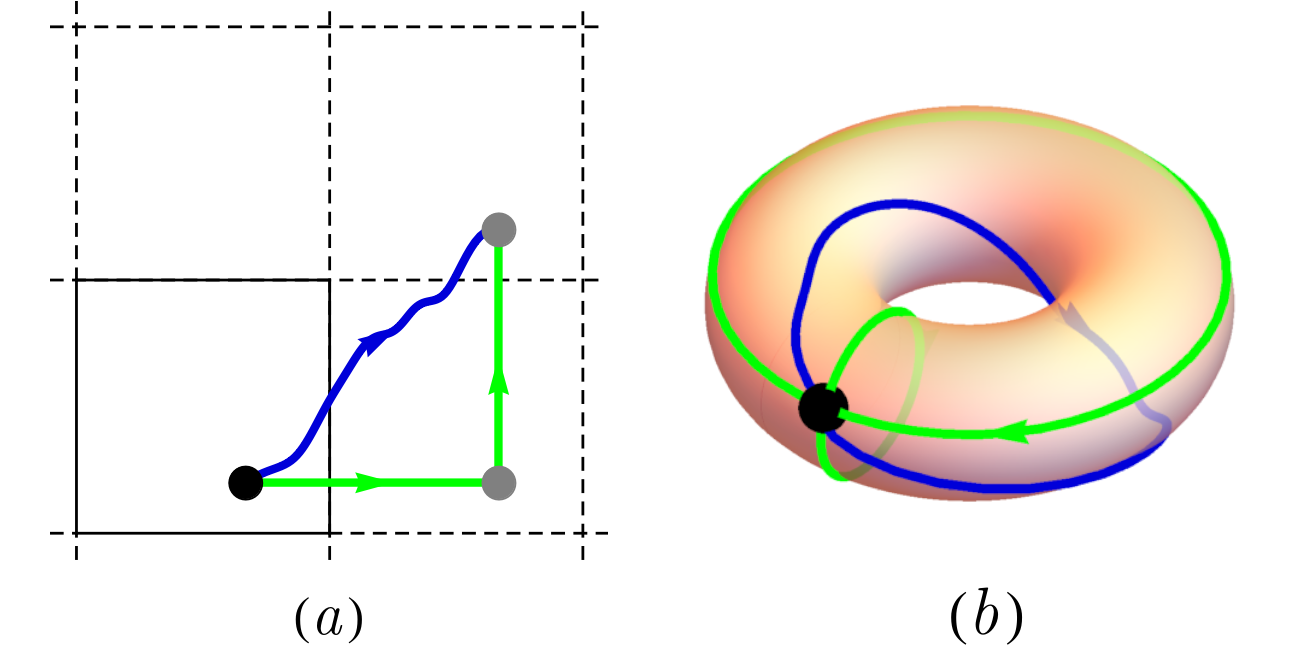}
    \caption{($a$)
    Two-dimensional atomic configuration space with periodic boundary condition. ($b$) Representation of the ACS on a 2-torus. Open paths in the plane whose end points are one the periodic image of the other map onto closed paths in the torus; closed paths in the plane map onto trivial loops in the torus (\ie closed paths that can be continuously shrunk to a point). The concatenation of the two green paths with the reversed blue one is a trivial loop on the torus. When strong adiabaticity holds, the total transported charge in a trivial path is zero.
    }\label{fig:additivity}
\end{figure}

Charge transport in ionic conductors occurs through the rearrangement of charge inhomogneities along with the motion of atomic nuclei in space. Ionic conductors are electronic insulators---a necessary condition for charge inhomogeneities to persist unscreened---so that their dynamics is accurately described in the adiabatic approximation, whereby electrons stay in their quantum ground state, while classical nuclei wander around in space. In the adiabatic approximation, the charge density depends on time through the dependence of the electronic Hamiltonian, $\hat H$, and its ground state, $\Psi_0$, on nuclear coordinates: $\hat H\bigl (\bm R(t) \bigr) \Psi_0(t)=E_0(t) \Psi_0(t)$, where ${\bm{R}=\{\bm{r}_1, \ldots,\bm{r}_N\}}$ is the set of positions of the $N$ atoms in the system. We call the space of all possible atomic configurations the \emph{atomic configuration space} (ACS), and its subspace whose configurations have a finite gap its \emph{adiabatic (sub-) space}. We describe macroscopic bodies using PBC with period $L$ along each Cartesian direction, as they are the only ones able to sustain a steady-state current in finite systems~\cite{pavarini2017physics}. When PBC are adopted, the ACS is isomorphic to a $3N$-dimensional torus and any path in ACS linking two periodic images of the same configuration is isomorphic to a \emph{closed path} on the torus. Paths in the adiabatic sub-space (be they closed or open) will be referred to as \emph{adiabatic paths}. On a torus, a closed path, $\mathcal C$, can be classified topologically by the number of windings it makes along each direction of the ACS ($n_{i\alpha}[\mathcal{C}]$, the \emph{winding numbers}): any such direction is identified by an atomic label, $i$, and by the Cartesian direction, $\alpha$, along which the atom moves. In the case depicted in Fig.~\ref{fig:additivity}, for instance, the horizontal green segment has winding numbers $(1,0)$, the vertical one $(0,1)$, and the blue path $(1,1)$. A closed path with all-zero winding numbers can be shrunk to a point on the torus and is called a \emph{trivial loop}.

The topological properties of the adiabatic sub-space will be key in our subsequent discussions, in view of which we introduce the concepts of \emph{adiabatic connectedness} and \emph{strong adiabaticity} (SA). The adiabatic sub-space is connected if any pair of points belonging to it can be joined by an (open) path entirely belonging to it; otherwise, it is the non-connected union of connected domains (in short: \emph{adiabatic domains}). An adiabatic domain is said to be \emph{strongly adiabatic} if any closed path belonging to it which is trivial on the torus is also trivial on it, \ie it can be continuously shrunk to a point without ever closing the gap. Equivalently, any two closed paths belonging to the same strongly adiabatic domain and featuring the same winding numbers can be deformed into one another without closing the gap ~\cite{foot_kernel}. This is obviously not the case if the paths belong to two disconnected strongly adiabatic domains.
In mathematical terms, SA amounts to saying that the \textit{fundamental group} of the adiabatic domain~\cite{fundamentalgroup} is a \textit{subgroup} of the fundamental group of the torus. Paths in the ACS can be parametrized by a fictitious time, so that the electronic Hamiltonian along a closed path is formally time-periodic. In a nutshell, the existence of closed paths that are trivial on the torus, but not in the adiabatic domain, entails that the time-dependent Hamiltonian evaluated along one these paths may sustain a Thouless' pump mechanism, with far-reaching consequences on the transport properties of the system, as we will demonstrate below.

The (fictitious) time parametrization of the electronic Hamiltonian along a closed adiabatic path, $\hat H(t+T) = \hat H(t)$, features a finite gap at all times. In his seminal paper on \emph{quantization of particle transport}~\cite{thouless83} Thouless showed that, in the adiabatic approximation, the dipole displaced during a time period (or, equivalently, along a closed adiabatic path $\mathcal{C}$), $\Delta\bm{\mu}(T)=\Delta\bm{\mu}[\mathcal{C}]$, is quantized in the large-$L$ limit:
\begin{equation}\label{eq: thouless theorem}
    \Delta\mu_\alpha[\mathcal{C}]=\oint_\mathcal{C} \dd{\mu_\alpha} =eLQ_\alpha[\mathcal{C}],
\end{equation}
where $ \dd\mu_\alpha = \frac{\partial \mu_\alpha(\bm{R})}{\partial \bm{R}} \cdot \dd\bm{R}$ is the electric dipole displaced along an infinitesimal segment of the path and $Q_\alpha[\mathcal{C}]$ is an integer-valued functional of the path, which is therefore constant for any continuous deformation of its argument within an adiabatic domain.

\begin{figure*}[t]
    \centering
    \includegraphics[width=1.7\columnwidth]{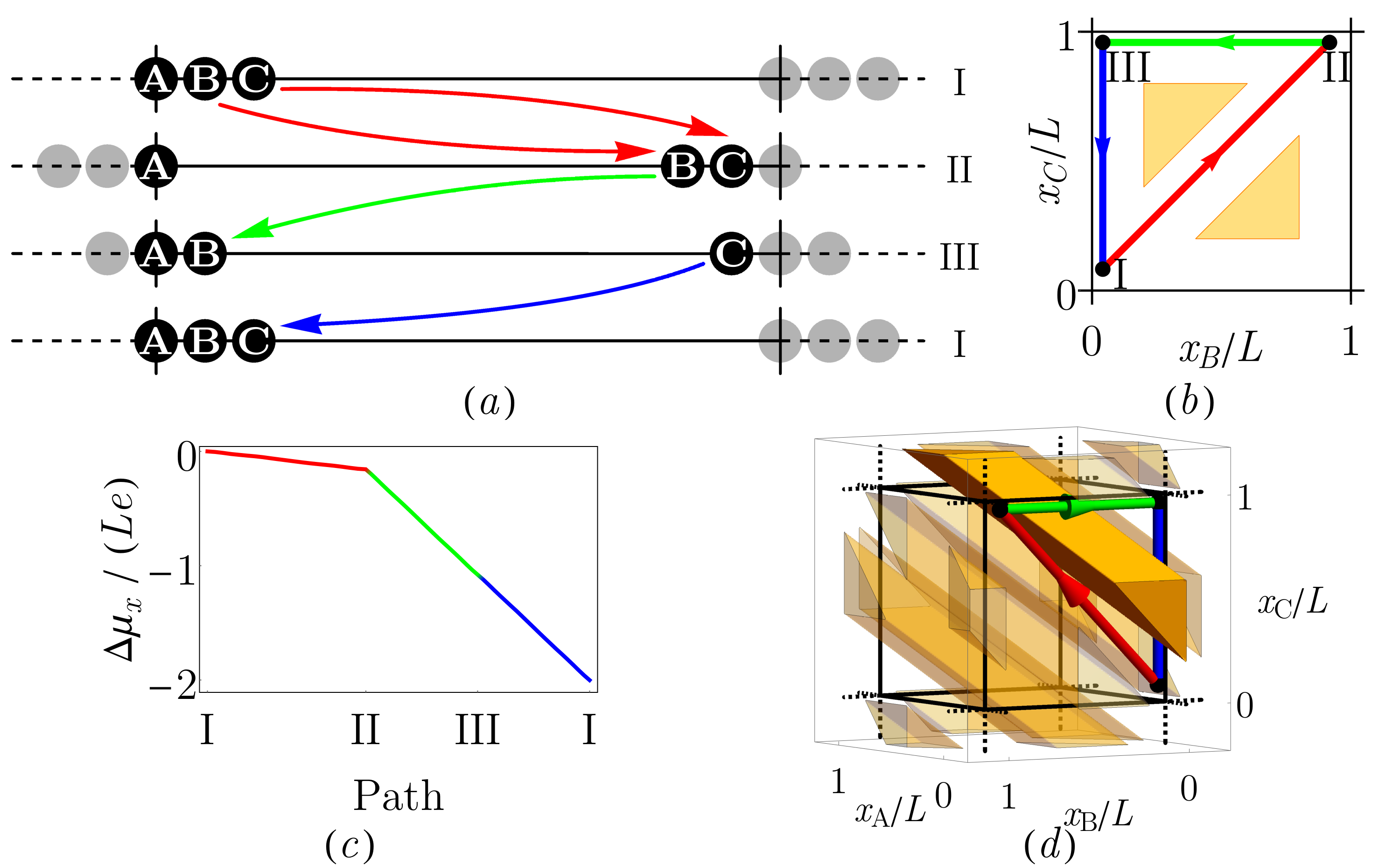}
    \caption{Linear H$_3^+$ in periodic boundary conditions. ($a$) I: Equilibrium configuration: the black circles and continuous line indicate atoms in the primitive supercell; gray circles and the dashed line indicate their periodic images. II and III are the two intermediate steps of the closed path described in the text. ($b$) Closed path in the 2D projection of the atomic configuration space relative to the atoms that participate in the loop; the yellowish areas indicate regions where the ground state is degenerate. ($c$) Dipole displaced along the closed path. Notice that the total displaced dipole is finite and an integer multiple of $eL$. ($d$) Closed path in the 3D projection of the ACS where also the $x$ coordinate of atom A is shown. The metallic region encircled by the path extends for all values of $x_A$ and the loop cannot be shrunk to a point without crossing it: the loop is thus non trivial. 3D projections onto subspaces where $x_\mathrm{A}$ is substituted with any other atomic coordinate have a similar appearance. An animation illustrating the trajectory of the atoms participating in the loop can be found in the file \texttt{Fig2\_animated.mp4} in the Ancillary Files. In the animation, the green dot indicates the position of the Wannier center of the two electrons in the system.}
    \label{fig: H3+ steps}
\end{figure*}

The properties of the charge $Q_\alpha[\mathcal{C}]$ strongly depend on the topology of the adiabatic domain to which $\mathcal{C}$ belongs. Since the charges associated to any pair of adiabatic paths that can be deformed into one another coincide, if this domain is strongly adiabatic we conclude that $Q_\alpha$ can only depend on the winding numbers of its argument: $Q_\alpha [\mathcal{C}] = Q_\alpha(\bm{n}_1, \bm{n}_2,\cdots \bm{n}_N)$, where $\bm{n}_i =(n_{ix}, n_{iy}, n_{iz})$ is the integer-valued 3-vector of the winding numbers of atom $i$. Furthermore, it must be \emph{additive} and \emph{isotropic}. Additivity means that $Q_\alpha$ is an integer-valued linear function of its integer arguments: $Q_\alpha (\{\bm{n}\}) = \sum_{i\beta} q_{i\alpha\beta} n_{i\beta}$; isotropy means that the $q$ coefficients are integer multiples of the identity: $q_{i\alpha\beta} = q_i \delta_{\alpha \beta}$. Additivity is illustrated in Fig.~\ref{fig:additivity}. The concatenation of the two green paths with the reversed blue one is a trivial path: if SA holds, the total charge transported along it must be zero, so that the charge transported along the blue path must equal that transported along the green ones. Isotropy follows from a slight generalization of this argument. The combination of additivity and isotropy allows one to identify the $q_i$ charge with the oxidation number of the $i$-th atom~\cite{grasselli2019}. When $N_S$ atoms of the same species $S$ are present, the ACS is the union of $N_S !$ domains that transform into each other under permutations. If these domains are strongly adiabatic and connected among themselves, one can swap two different atoms of the $S$ species without closing the electronic gap, implying that all the atoms of that species have the same OS. When two permutational domains are not connected, but still strongly adiabatic, an OS can still be assigned to each atom within a same domain, but atoms of the same chemical species belonging to different domains may feature different OSs: this is the case, \eg, of ferrous-ferric water solutions. When SA does not hold, instead, charge transport can no longer be characterized in terms of winding numbers on the torus, and the very concept of OS looses much of its topological meaning, thus opening the way to a transport regime where charge can be moved without any concomitant atomic displacements.

Let us conclude this theory section with a summary of the links between transport, topology, and Thouless' pumps. The key concept here is that PBC endow the ACS with the topology of a $3N$-torus. Closed paths (loops) on the torus are classified according to their winding numbers, $\bm n\in \Z^{3N}$: loops with $\bm n=0$ represent \emph{non-convective} trajectories, \emph{i.e.} trajectories where each atom returns to its original position; loops with $\bm n \neq 0$, instead, describe trajectories where one or more atoms move from the initial position to one of their periodic images. Of course, genuine convective trajectories are not loops (they are almost surely open paths), but any of them can still be decomposed into a (possibly non-trivial) loop concatenated with an open path whose length is bounded by a quantity independent of $\bm n$: this allows one to express the large-time diffusive behavior of atomic trajectories---and hence transport properties---in terms of the topology of the system's adiabatic subspace~\cite{grasselli2019}. The next important concept is that, along a loop, the electronic Hamiltonian is formally cyclic. Thouless' quantization of particle transport~\cite{thouless83} can then be invoked to infer that the dipole displaced along a loop is in general non-vanishing and quantized. If SA holds, atomic OSs can be rigorously defined as topological charges~\cite{grasselli2019,rappe2012}. Under this condition, the electric dipole displaced along an open trajectory is, up to a term that is negligible in the large-time limit, a linear combination of the winding numbers of an aptly defined loop, the atomic OSs being the coefficients. As the winding numbers are a measure of the overall atomic displacements, we conclude that charge transport is necessarily convective in this case. Note that, even under SA, the dipole displaced along an $\bm n\ne 0$ loop in ACS is itself induced by a trivial Thouless' pump: one that is simply driven by the windings along each direction of the torus. For this reason, the transport regime described by this mechanism is aptly dubbed \emph{trivial}. Non-trivial transport may occur whenever in an adiabatic domain there exist loops with $\bm n = 0$ that cannot be shrunk to a point without crossing a zero-gap region, \ie when SA is broken. Thouless' theory can then be leveraged again to conclude that the dipole displaced along any such loop is quantized and possibly non-zero, whereas the atomic displacements vanish because so do the winding numbers. This is a regime where the same topological features of the ACS that would give rise to a non-trivial Thouless' pump mechanism along closed paths, also give rise to non-convective charge transport, when applied to open paths.

\begin{figure*}[t]
    \centering
    \includegraphics[width=1.7\columnwidth]{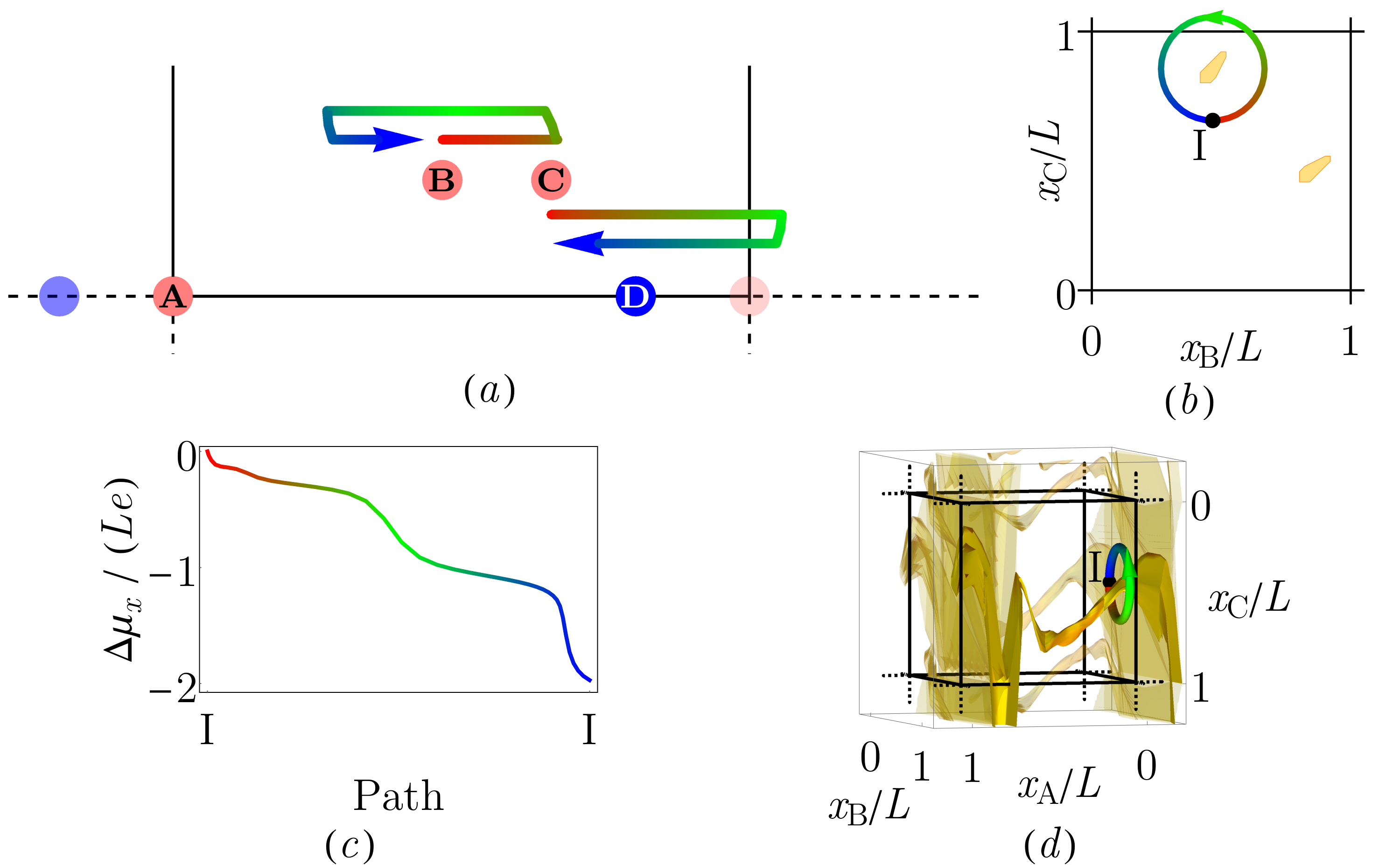}
    \caption{
    A planar configuration of the $\mathrm{K_3Cl}$ system undergoing a loop in atomic configuration space. ($a$) Initial and final configurations. K and Cl atoms are indicated by pink and blue circles, respectively. The colored curved arrows indicate the 1D trajectories of the two K atoms participating in the loop. The color encodes the fictitious time parametrizing the loop ($\mathrm{red\to blue}$). ($b$) Closed path in the 2D projection of the atomic configuration space  relative  to  the  atoms  that  participate  in  the  loop; the yellowish areas indicate regions where the ground state is degenerate. ($c$) Dipole displaced along the closed path. ($d$) Closed path in the 3D projection of the ACS where also the $x$ coordinate of atom A is shown. An animation illustrating the trajectory of the atoms participating in the loop can be found in the file \texttt{Fig3\_animated.mp4} in the Ancillary Files. The green dots indicate the positions of the Wannier centers of the electrons in the system.}
    \label{fig:K3Cl}
\end{figure*}

\section{Non-trivial transport}
The simplest system displaying a non-trivial Thouless' pump mechanism is probably the (linear) tri-hydrogen cation, $\mathrm{H_3^+}$: three protons are aligned and the ground state of the two electrons is a singlet, therefore non-degenerate, resulting in a theoretical equilibrium interatomic distance ${\Delta=0.826\,\mathrm{\AA}}$. We treat the molecule using PBC with period ${L=10.6\,\mathrm{\AA}\gg\Delta}$ along the three Cartesian directions, thus amounting to enclose the molecule in a cubic supercell of side $L$. Further computational details are given in Appendix~\ref{App:computational}. We now consider a closed path in ACS, consisting of the following three steps (see Fig.~\ref{fig: H3+ steps}):
\begin{enumerate}[noitemsep,nolistsep,leftmargin=*]
    \item The B and C protons are first rigidly translated towards the end of the supercell until the distance between C and the periodic image of A (located a $x=L$) is $\Delta$ (red arrows, ending at configuration II);
    \item The B proton, now located at $x=L-2\Delta$, is moved back to its original position (green arrows, ending at configuration III);
    \item Finally, the C proton is moved back from $x=L-\Delta$ to its original position (blue arrows, ending at configuration I).
\end{enumerate}
This path is periodic (\emph{i.e.} it is isomorphic to a trivial path on the torus), the last configuration being equal to the first. The ionization potential of $\mathrm{H}_2$ is larger than that of $\mathrm H$; therefore, along the path the pair of protons that is displaced or left behind stays neutral, and the electronic gap is always larger than that of $\mathrm{H}_2$. The ground state remains a singlet throughout the entire trajectory, as we explicitely checked by computing the total energy of the system for both the singlet and the triplet spin states. If SA holds, the charge transported along the path must vanish trivially; otherwise, a Thouless' pump mechanism may allow a non-trivial charge transport. In order to check if the latter case occurs, we have computed the total dipole displaced along each segment of the path, $\Delta\bm{\mu}$, according to the modern theory of the polarization in the Wannier representation~\cite{vanderbilt2018berry},
%
\begin{equation}
    \Delta\bm{\mu}_{IF} = e\int_{I}^{F} \left(\sum_{i=1}^{N} Z_i \dd{\bm{r}_i}  -2 \sum_{j=1}^{M} \dd{\bm{w}}_{j} \right), \label{eq:core + tutti WC}
\end{equation}
%
where $Z_i$ is the positive core charge of atom $i$ ($Z_i=1$, in the present case), $\bm{w}_{j}$ the position of the Wannier Center (WC) associated to the $j$-th occupied electronic band of the system, $M$ is the number of occupied states ($M=1$, in the present case), and the factor 2 in front of the second sum accounts for the double occupancy of each molecular orbital. Our results, displayed in Fig.~\ref{fig: H3+ steps}$(c)$, indicate that a net charge, $Q=-2e$ is displaced along the path, thus revealing the existence of non-adiabatic domains in the ACS that the path loops around. Indeed, when the distance between any pair of protons is much larger than the molecular bond length, the ground state consists of two neutral atoms and one proton, and it is degenerate, because it does not matter which atoms are neutral and which one is ionized. The regions where this condition occurs is highlighted with yellowish triangles in Fig.~\ref{fig: H3+ steps}$(b)$, revealing that it is in fact encircled by the closed path. When the full 9-dimensional ACS is considered, the plane depicted in  Fig.~\ref{fig: H3+ steps}$(b)$ is the locus where all the coordinates vanish but $x_\mathrm{B}$ and $x_\mathrm{C}$, and the triangles are the bi-dimensional sections over this plane of hyperprisms that pierce the entire ACS so that the loop cannot be shrunk to a point without closing the gap even when embedded in the full 9-dimensional space (\ie the loop is \textit{non trivial} in the adiabatic subspace), as illustrated in Fig.~\ref{fig: H3+ steps}($d$).

Note that, while the total dipole is ill defined (both because it is intrinsically so when computed in PBC~\cite{martinPBC}, and because the system is charged), dipole differences are perfectly well defined also in this case. Our previous considerations on the relative magnitude of the ionization potentials of atomic and molecular hydrogen imply that WCs move (almost) rigidly with the proton \emph{pair} being displaced, as illustrated in the animation \texttt{Fig2\_animated.mp4} to be found in the Ancillary Files. This implies that, when displaced individually, protons carry a unit charge, and one would be tempted to attribute an OS $q_\mathrm{H}=1$ to each of them. However, when they move in pairs, they carry a zero charge, a manifest breakdown of charge additivity, due to the breakdown of strong adiabaticity. The overall effect of the different charges transported by H atoms according to whether they are displaced individually or in pairs is that the total charge transported along the closed path of Fig.~\ref{fig: H3+ steps} does not vanish, while the net mass does.

The existence of adiabatic transport anomalies entails the occurrence of two partially conflicting requirements: a high degree of ionicity and the presence of loosely bound localized electron states that can wander through the system without ever closing the gap. Non-stoichiometric molten salts seem therefore ideal candidates to display non-trivial transport~\cite{selloni1987, fois1988bipolarons}. In order to prepare for the study of such systems, we examine now the simplest molecular system possibly displaying their essential electronic features: the neutral $\mathrm{K_3Cl}$ complex. Computational details can be found in Appendix~\ref{App:computational}. In Fig.~\ref{fig:K3Cl} we show a planar configuration of this system along with a closed path in ACS displaying charge transport without any net mass displacement. As in the previous case, the ground state of the system is a singlet throughout the whole trajectory. The dipole displaced by moving each of the atoms to their periodic images in the neighboring cell along the $x$ direction is, in units of $e L$, equal to +1 for A and C, equal to -1 for B and D. Conversely, moving any of the atoms to its periodic image in a direction perpendicular to the $x$ axis would break the bond of such atom with the rest of the system, resulting in a degenerate ground state. Moreover, as we verified that it is possible to swap atoms B and C without closing the gap, there is no way to uniquely associate an integer charge to each atom, whose OS would thus be topologically ill-defined. Based on our previous arguments, we conclude therefore that SA is violated here again. In fact, we identified a region in the ACS where the gap closes, as indicated by the shaded area in Fig.~\ref{fig:K3Cl}$(b)$, and we considered an adiabatic closed path in the ACS which is trivial on the torus, but loops around such region. The $x$ component of the electric dipole displaced along this loop is plotted in Fig.~\ref{fig:K3Cl}$(c)$. While the total charge displaced along the $y$ and $z$ directions vanishes, we observe an integer charge equal to $-2e$ pumped along the $x$ direction. Here again, we see that an integer charge is adiabatically transported without any net atomic displacements. This Thouless' charge-pump mechanism can be entirely ascribed to the loosely bound highest-occupied molecular orbital (HOMO). In fact, we checked that the contribution of the lower-lying molecular orbitals to the dipole displaced along the loop vanishes. By the same token, if one computes the charges associated to individual atomic displacements as above, but only considering the contributions to the displaced dipole from lower-lying molecular orbitals, we obtain $+1$ for all the K atoms and $-1$ for Cl. This is best illustrated by following the motion of the various WCs displayed in the animation quoted in the caption of Fig.~\ref{fig:K3Cl}.

\begin{figure}
    \centering
    \vspace{5mm}
    \includegraphics[width=0.9\columnwidth]{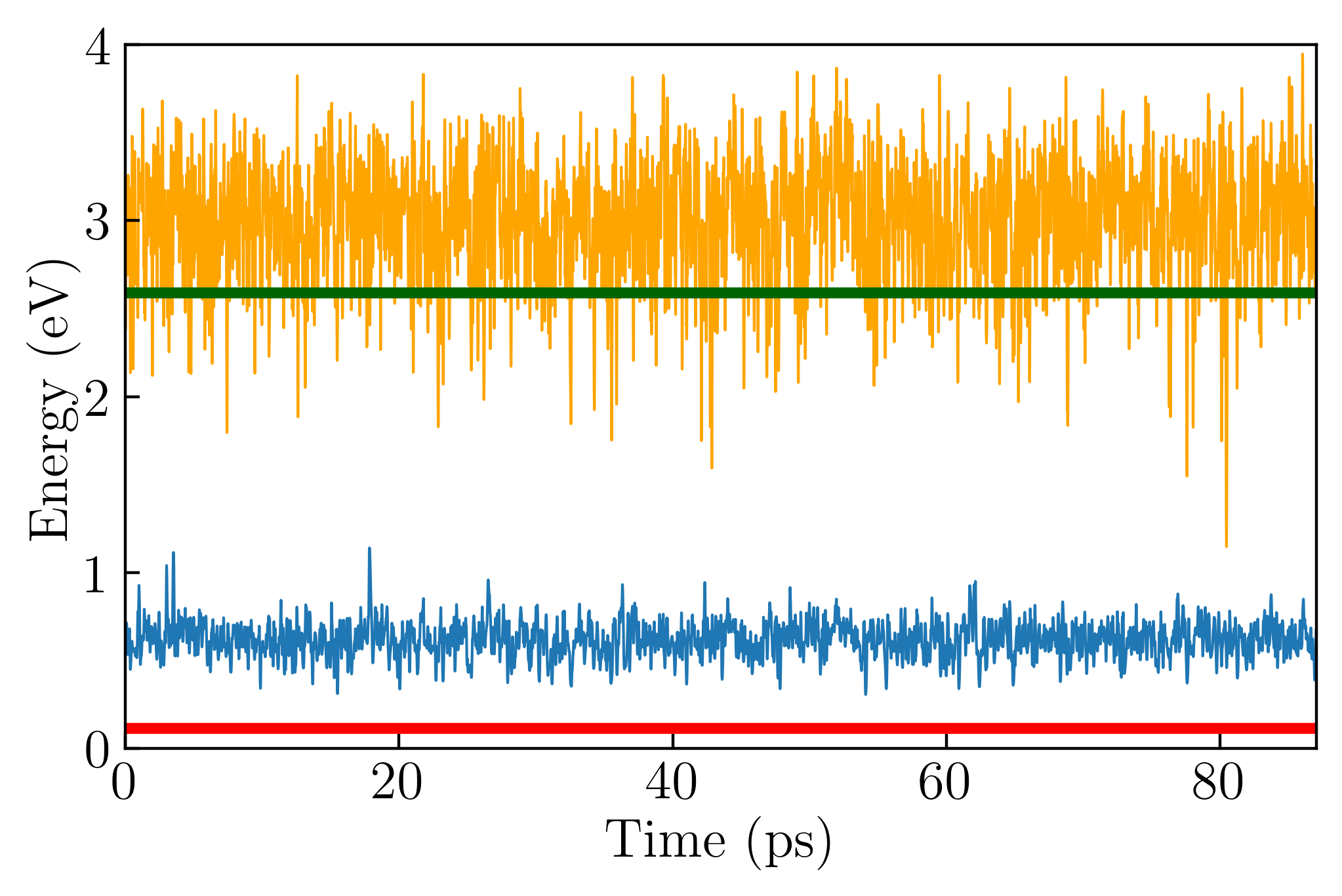}
    \caption{
        Time series of the HOMO/LUMO (blue) and HOMO--1/LUMO (orange) energy gaps. The horizontal red line indicates the thermal energy, $\kb T$. The horizontal green line is the average HOMO/LUMO gap for the stoichiometric $\mathrm{K_{32}Cl_{32}}$ system. }\label{fig: gap}
\end{figure}

We now move to a more realistic system and consider a model for a  ${\mathrm{K}_{x}(\mathrm{KCl})_{1-x}}$ dilute liquid metal/metal-halide solution. The equilibrium properties of non-stoichiometric molten salts are characterized by the existence of localized electron pairs, often referred to as \emph{bi-polarons} \cite{selloni1987,selloni1987electron,fois1988bipolarons}. The formation of bi-polarons is made possible by the balance between the increase in the quantum kinetic and electrostatic electron-electron repulsion energies resulting from the localization of the solvated electron pair and the attractive electron-cation energy gained by accommodating the pair into a cationic hollow, often described as a liquid-state analog of an F center in a crystal; charge transport can then be assimilated to the fast diffusion of the solvated electrons followed by their temporary stabilization in cationic hollows, driven by thermal fluctuations \cite{selloni1987,selloni1987electron,fois1988bipolarons}. The very existence of such an adiabatic hopping-like mechanism breaches the compelling topological constraints that SA sets on ionic conduction \cite{grasselli2019} and could not be possible without breaking the latter.

We model the melt with 33 K atoms and 31 Cl atoms, corresponding to a concentration ${x \approx 0.06}$.
This model can be qualitatively described as made of 31 $\mathrm{Cl}^-$ anions and 33 $\mathrm K^+$ cations, with the addition of two neutralizing solvated electrons whose dynamics is only weakly correlated with the ionic motion \cite{selloni1987,selloni1987electron,fois1988bipolarons}. 

We simulate this system within density-functional theory (DFT) using Car-Parrinello \emph{ab initio} molecular dynamics (AIMD)~\cite{Car1985}. Our simulations are performed using a cubic supercell with side ${L=14.07\,\mathrm{\AA}}$, corresponding to a density ${\rho=1.42\,\mathrm{g/cm^3}}$ at a temperature of ${T = (1341\pm 93) \, \mathrm{K}}$ (the incertitude on the value of the temperature is a finite-size effect, whereas the statistical incertitude on the average is two orders of magnitude smaller). Further technical details are given in Appendix~\ref{App:computational}. The dynamics of the system is restricted to the singlet energy surface, as we explicitly verified that the triplet one consistently lies $\approx 0.40\un{eV}$ above. This being the case, the system is closed-shell. Nonetheless, the presence of unpaired solvated electrons would not affect our conclusions on non-trivial transport, as long as each spin channel stays gapped and dynamically decoupled from the other, and the system's electronic insulating character and adiabatic evolution are preserved~\cite{selloni1987,RestaSorella1999}. In Fig.~\ref{fig: gap} we display the time series of the energy gap between the HOMO and the \emph{lowest-unoccupied molecular orbital} (LUMO), as well between the molecular orbital just below the HOMO (HOMO--1) and the LUMO. The HOMO--1 corresponds to the highest molecular orbital localized on Cl$^-$ anions, whereas the HOMO is occupied by the solvated lone pair (see below). The numerical values of these energies are affected by DFT errors that lead to an underestimate of the electronic gaps. Notwithstanding, the system stays electronically insulating all along the AIMD trajectory, thus confirming the adequacy of an adiabatic treatment of transport in these systems for small enough concentrations. The average HOMO--1/LUMO gap would coincide with the average stoichiometric HOMO/LUMO gap for extremely low concentrations ($x\to0$). In this limit, the energy level of the lone pair (\emph{i.e.} the HOMO of the non-stoichiometric system) corresponds to a donor impurity level in the HOMO/LUMO gap of the stoichiometric system, slightly below the bottom of the empty-state band. As the concentration increases, the impurity level broadens to a band,
which eventually merges into the empty-state band of the stoichiometric system, thus turning the electrolyte into a metal. If the states near the Fermi energy stay localized by disorder, this transition would be delayed until the Fermi energy crosses the mobility edge. In either case, we believe that our conclusions hold in the electrolyte regime. The value of the concentration appropriate to our model system, $x\approx 0.06$, is below the critical value $x_c\approx 0.1$, at which an insulator-to-metal transition is expected to occur, resulting from experimental~\cite{warren1987metal} and numerical~\cite{fois1989approach, metaltransition-frenkel} evidence. A similar behavior has been recently evinced using photo-emission spectroscopy for alkali-metal solutions in liquid ammonia~\cite{buttersack2020photoelectron}.

\begin{figure}[hb!]
    \centering
    \includegraphics[width=0.8\columnwidth]{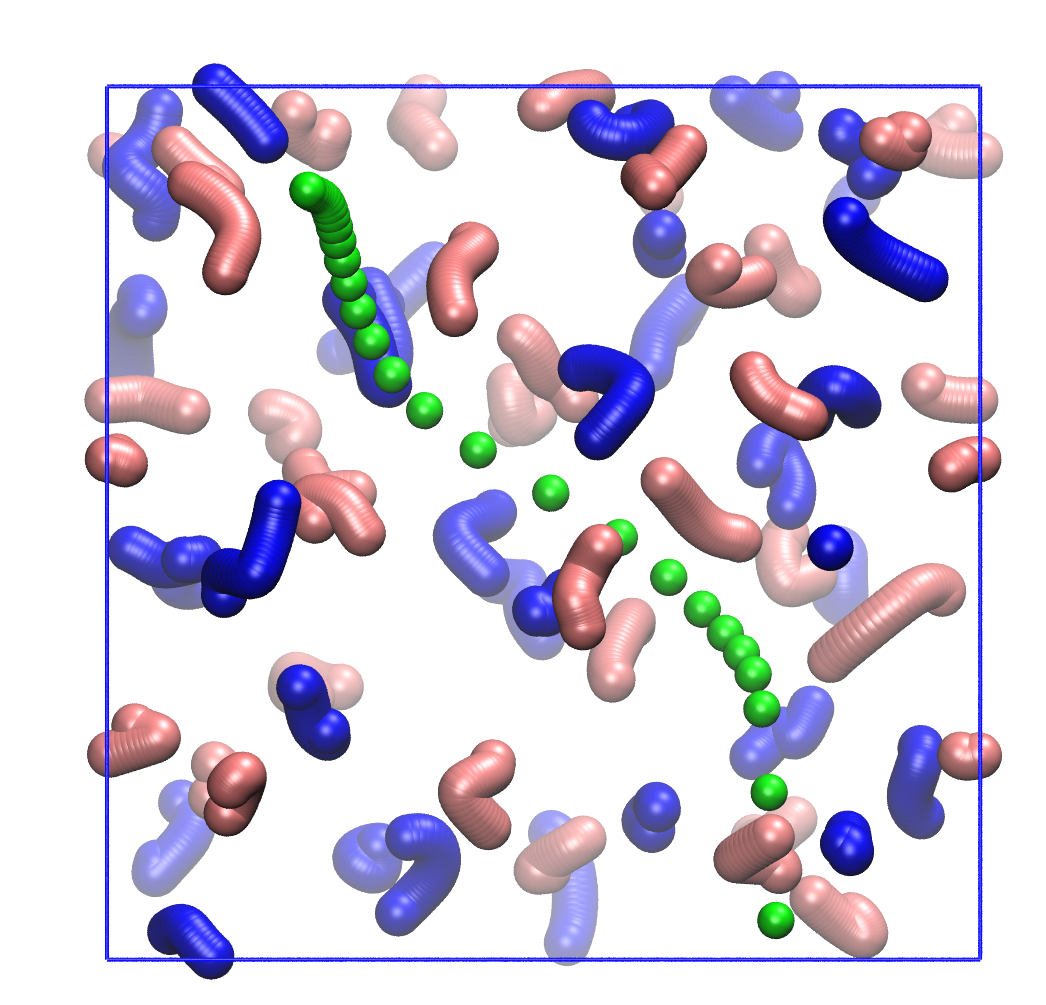}
    \caption{
        Overlay of  several  consecutive  snapshots from a $435 \un{fs}$-long sample of AIMD trajectory of our $\mathrm{K_{33} Cl_{31}}$ model. K$^+$ ions are depicted in pink, Cl$^-$ ions in blue, while the Wannier center associated to the lone bi-polaronic pair is displayed in green.
    }\label{fig:vmd scene}
\end{figure}

\begin{figure*}[t]
    \centering
    \includegraphics[height=0.4\textheight]{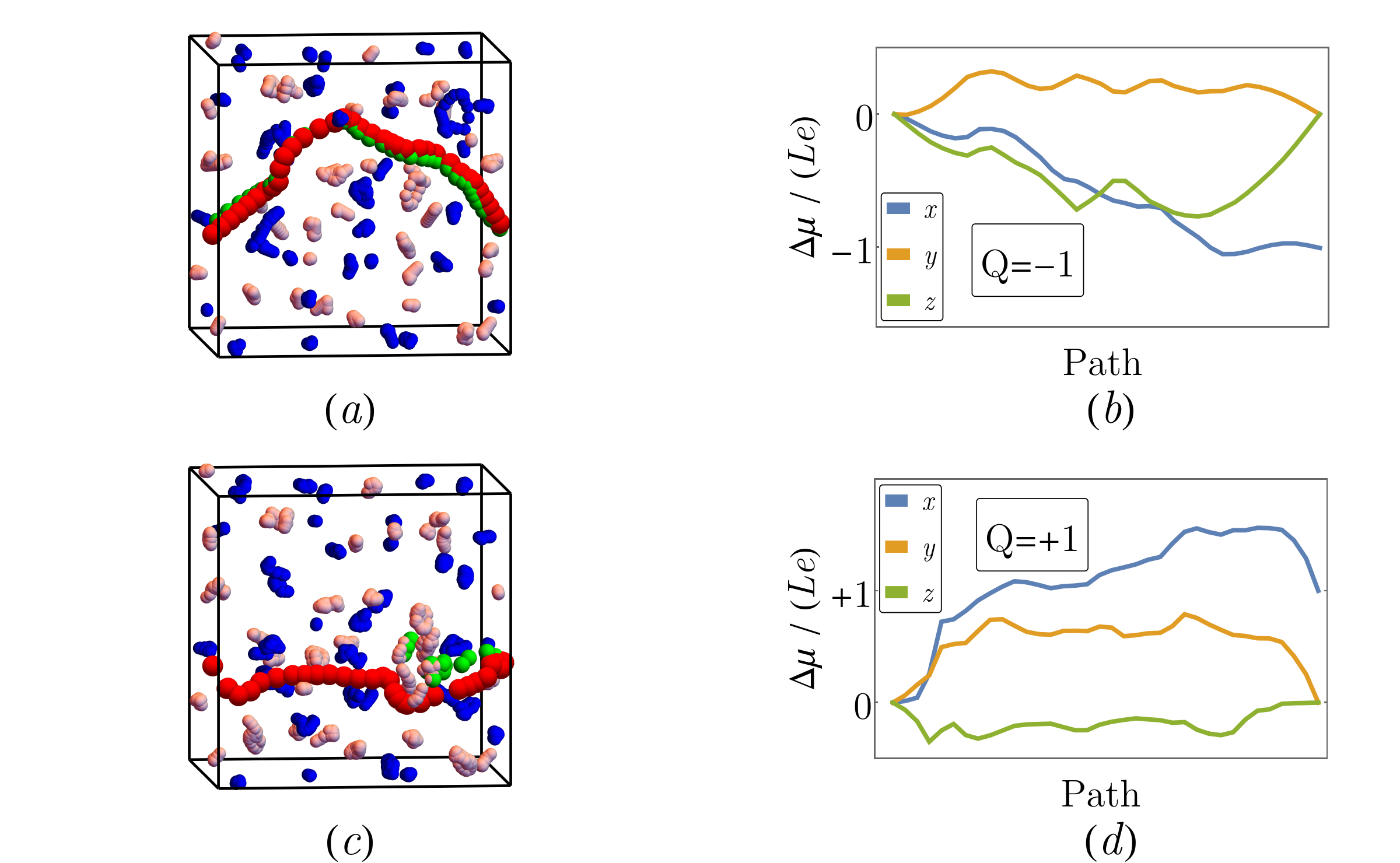}
    \caption{
        ($a$) and ($c$): Different loops in the $\mathrm{K_{33}Cl_{31}}$ atomic configuration space, whose initial and final configurations are the same. Cl atoms are depicted in blue. All K atoms but one are depicted in pink. One selected K atom, depicted in red, is moved from its initial position to its periodic image along the $x$ direction, thus featuring a winding number $n_{ix}=+1$. All the other atoms feature zero winding numbers. The position of the lone pair is depicted in green. ($b$) and ($d$): Dipoles displaced along the closed paths depicted on their left. The charge displaced along the two paths differ, in spite of the displacement of the same K atom and the same winding numbers. This indicates that no oxidation state can be uniquely associated to the (arbitrarily) chosen K atom, and transport anomalies have to be expected.
    } \label{fig:vermi}
\end{figure*}

In Fig.~\ref{fig:vmd scene} we display the overlay of several consecutive snapshots from a short segment of our AIMD trajectory. One sees that, by the time the lone pair has covered a distance comparable to the size of the supercell, all the atoms have traveled only a small fraction of this length. This suggests that charge transport in these systems may be strongly affected by the dynamics of the localized lone pairs, whose very existence we have seen to be closely related the topological properties of the electronic ground state. The animation contained in the file \texttt{K-KCl-trj.mp4}, to be found in the Ancillary Files, confirms that the lone pair diffuses much faster than the atoms (the color code is the same as in Fig.~\ref{fig:vmd scene}). Its motion, while being uniquely determined by the atomic  adiabatic dynamics, is largely uncorrelated from it. It can thus give rise to a non trivial transport regime such that electric currents are mainly carried by solvated electrons, not corresponding to any atomic displacements, which is essentially made possible by the breaking of SA. To show this, we computed the dipole displaced along two properly designed loops in the $\mathrm{K_{33}Cl_{31}}$ ACS, beginning and ending at the same configuration, where one K atom is moved from its initial position to one of its adjacent periodic images along the $x$ axis, as depicted in Fig.~\ref{fig:vermi}. The two loops have identical winding numbers: ${\bm{n}_\mathrm{K} = (1,0,0)}$ for the moving K, and ${\bm n = (0,0,0)}$ for all the other atoms. Nonetheless, the dipoles displaced along them differ, as reported in panels $(b)$ and $(d)$, corresponding to two different topological charges ($Q=\pm 1$) for the same K atom. Such a state of affairs is a clear evidence that the two loops cannot be deformed into one another without hitting a non-adiabatic region, thus making it impossible to assign a well defined OS to each atom using the procedure of Refs.~\onlinecite{rappe2012,grasselli2019}, in striking contrast with chemical common sense. Even though in a physical trajectory no such loops in the ACS are expected to occur, nor will a lone pair stay bound to the same ion for much longer than a fraction of the atomic diffusion time, this thought experiment clarifies the links between SA breaking and the establishment of a regime where loops in ACS can be described by non-trivial Thouless' pumps and open trajectories may carry a charge current not corresponding to any ionic currents.

\begin{figure}
    \centering
    \includegraphics[width=0.9\columnwidth]{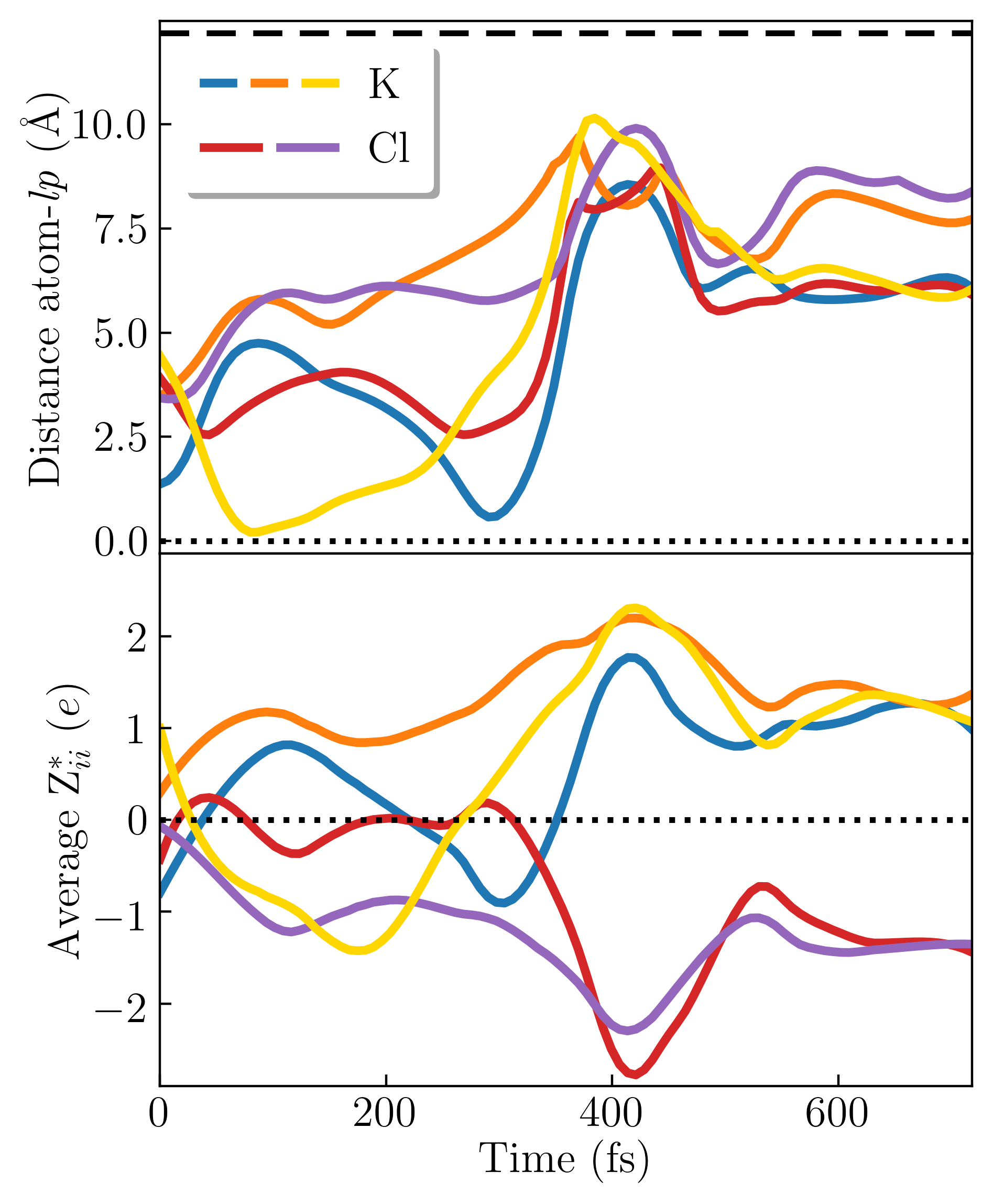}
    \caption{Upper panel: time series of the distances from the electron lone pair in K$_{33}$Cl$_{31}$ of the five nearest atoms. The horizontal lines are guides for the eye: dotted, distance equal to zero; dashed, maximum distance allowed in PBC, \emph{i.e.} $\sqrt{3}L/2$. Lower panel: average diagonal elements of the Born effective-charge tensor of the five atoms described above; the horizontal dotted line marks the zero value.}
    \label{fig:zborn_evolution}
\end{figure}

The transport anomalies displayed in Fig.~\ref{fig:vermi} are expected to show conspicuously in the behavior of the ionic Born effective charges, in terms of which the instantaneous charge current can be expressed~\cite{grasselli2019}. The animation contained in the file \texttt{zborn.dynamics.mp4}, to be found in the Ancillary Files, displays a typical segment of an AIMD trajectory. The upper panel of Fig.~\ref{fig:zborn_evolution} displays the time series of the distances from the lone pair of the five atoms originally closest to it. The lower panel displays the corresponding average diagonal elements of the effective-charge tensors. One sees that, when all the atoms are stably far from the lone pair ${(t\gtrsim 500 \un{fs})}$, the values of the effective charges are close to what chemical intuition would suggest ${(\mathrm{Z}^*_\mathrm{K}\approx +1}$ and ${\mathrm{Z}^*_\mathrm{Cl}\approx -1)}$. When some of them approach the lone pair, instead, weird things may occur. For one thing, the effective charge of the K atom closest to the lone pair may go negative, meaning that the latter is provisionally dragged by the K ion along its movement. For a second, the effective charge of a Cl atom, while never quite positive, may nearly vanish when it gets close enough to the lone pair; this is likely due to the screening effect of the pair's highly polarizable wavefunction. For a third, effective charges change abruptly when passing from an anomalous to a normal regime: the K effective charge may become relatively large and positive just after having gone negative, and a few Cl charges may correspondingly become relatively large, while staying negative, so as to preserve local charge neutrality. The duration of this transition, a few dozen femtoseconds, is the time it takes for the lone pair to abruptly change its local environment, as witnessed by the steep change of the distances from it of the atoms considered in Fig.~\ref{fig:zborn_evolution}.

\begin{figure}
    \centering
    \includegraphics[width = 0.9\columnwidth]{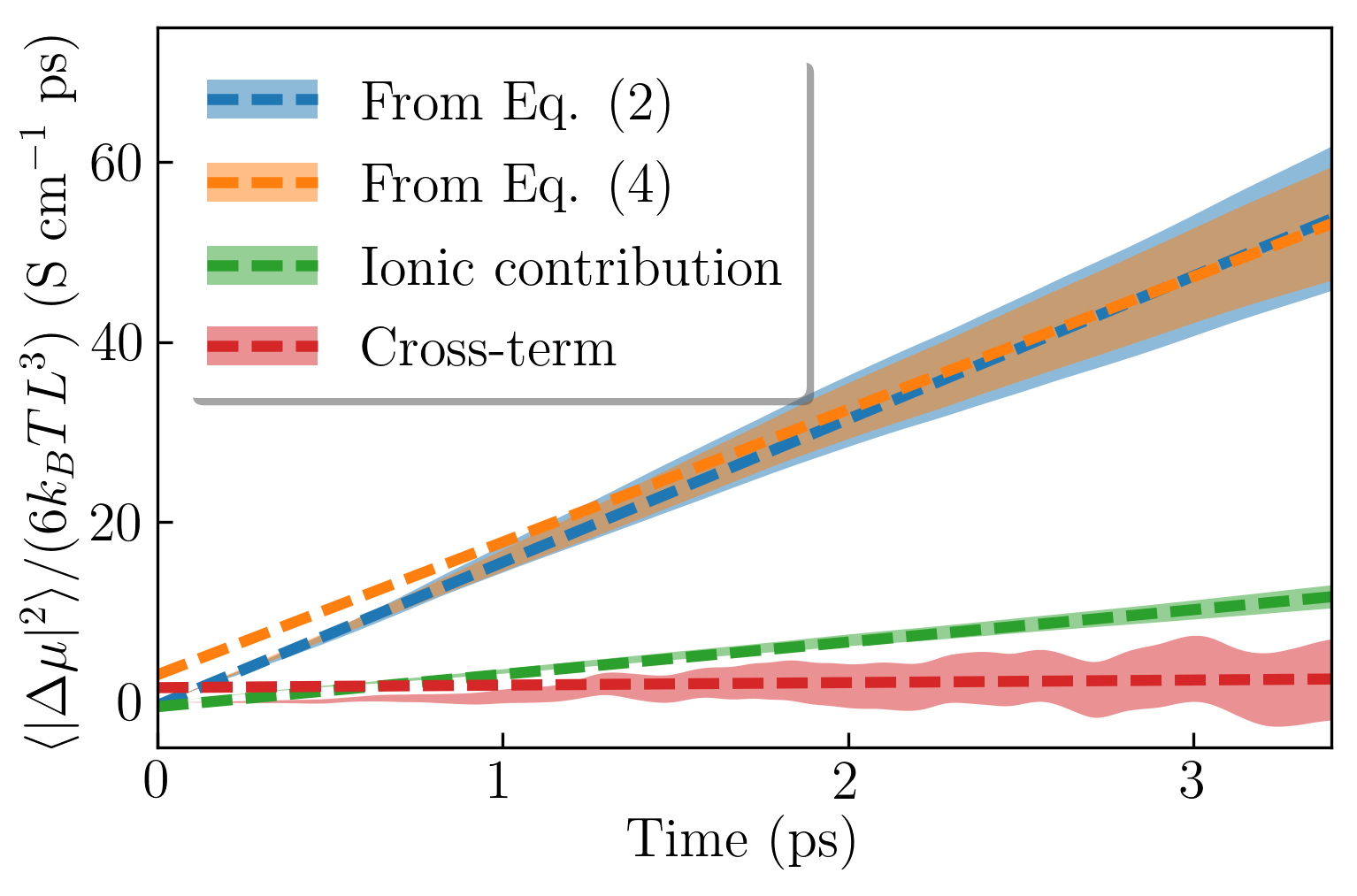}
    \caption{Adiabatic charge transport in molten $\mathrm K_{0.06}/\mathrm{(KCl)}_{0.94}$. Time series of the mean-square displaced dipole from definitions~\eqref{eq:core + tutti WC} (blue) and~\eqref{eq:no + lb WC} (orange). The contribution due to the ionic cores and the tightly bound electrons is shown in green. The cross-correlation term is depicted in red. According to Eq.~\eqref{eq:sigma Einstein}, the slope of the straight lines is a measure of the electric conductivity, whose actual value is estimated from cepstral analysis, as explained in Appendix~\ref{App:cepstral}.}
    \label{fig:K-KCl fuso}
\end{figure}

In order to evaluate the impact of non-trivial transport on the electrical conductivity of our system, we computed it using the Helfand-Einstein relation~\cite{Helfand1960,grasselli2019},
\begin{equation} \label{eq:sigma Einstein}
    \sigma = \frac{1}{3 L^3 \kb T} \lim_{t\to\infty} \frac{\langle \left| \Delta \bm{\mu}(t) \right|^2 \rangle }{2t},
\end{equation}
where $\Delta\bm{\mu}(t)$ is the electric dipole displaced along the AIMD trajectory in a time $t$. $\Delta\bm{\mu}(t)$ has been alternatively computed from Eq.~\eqref{eq:core + tutti WC} and from:
\begin{equation}\label{eq:no + lb WC}
        \Delta \bm{\mu}^*_{IF} = e \int_I^F \left( \sum_i q_i \, \dd\bm{r}_i - 2 \dd \bm{w}_{\lb} \right)
\end{equation}
where $\bm{w}_{\lb}$ is the position of the lone-pair WC, $q_i=+1$ for K atoms, $-1$ for Cl atoms, and the factor $q_{\lb}=-2$ reflects the occupancy of the loosely bound HOMO. The definition of $\Delta \bm{\mu}^*_{IF}$ differs from that of $\Delta \bm{\mu}_{IF}$,  Eq.~\eqref{eq:core + tutti WC}, in that in Eq.~\eqref{eq:no + lb WC} a fixed oxidation state is associated with all the atoms of a same species in the spirit of Ref.~\onlinecite{grasselli2019}---and as it would be in a stoichiometric mixture---while the lone pair occupying the localized and loosely bound HOMO is treated as an independent charge carrier.
Our results, illustrated in Fig.~\ref{fig:K-KCl fuso}, yield the values ${16.2 \pm 0.8 \un{S/cm}}$ and ${15.9 \pm 0.8 \un{S/cm}}$ for the conductivities computed from definitions~\eqref{eq:core + tutti WC} and~\eqref{eq:no + lb WC}, respectively \nocite{c_foot,ercole2017accurate,bertossa}~\cite{c_foot}. All the numerical values of the transport coefficients reported here have been evaluated using the \emph{cepstral analysis} method~\cite{ercole2017accurate,bertossa}, as briefly explained in Appendix~\ref{App:cepstral}. We see that the two values coincide within statistical errors, giving substance to our topological analysis of non-trivial transport in these systems. Maybe fortuitously, these values compare well with the experimental data at such concentration of K atoms \nocite{d_foot,nattland1986experiment,bronstein1958electrical}\cite{d_foot}.
The conductivity is much larger than the value obtained from the ionic contribution in Eq.~\eqref{eq:no + lb WC} ($3.6 \pm 0.3\,\mathrm{S/cm}$, the green line in Fig.~\ref{fig:K-KCl fuso}): this indicates that the conductivity is almost entirely determined by the diffusion of the solvated lone pair, and is in fact much larger than it typically is in stoichiometric molten salts (\ie, $ 3.2\pm0.2\un{S/cm}$~\cite{grasselli2019}). Furthermore, we observe that the total conductivity coincides with the sum of the ionic and lone-pair contributions, implying that the cross correlation resulting from the product of the first and second terms on the right-hand side of Eq.~\eqref{eq:no + lb WC} is negligible, as confirmed by the vanishing slope of the red curve in Fig.~\ref{fig:K-KCl fuso}. We computed the diffusivity $D_i$ of each species ${i=\mathrm{K},\,\mathrm{Cl},\,lp}$ according to the Einstein formula:
\begin{equation}
    D_i = \lim_{t\to\infty}\frac{\left<|\Delta \bm{r}_i(t)|^2\right>}{6 t}.
\end{equation}
The mobilities, ${\mu_i\equiv q_i e D_i/(\kb T)}$, are then estimated to be $1.23 \pm 0.02$, $1.14 \pm 0.07$, and $102 \pm 5$ $(10^{-3}\un{cm^2\,V^{-1}\,s^{-1}})$ for K, Cl, and the lone solvated pair, respectively. The lone-pair mobility is two orders of magnitudes larger than the ionic ones, in agreement with experimental evidence~\cite{warren1987picosecond} and with the observed predominance of the lone-pair contribution to the total conductivity.
\section{Conclusions}
Conducting materials are usually classified into two broad families: metals and ionic conductors. The charge carriers of metals are electrons, whose equilibrium and dynamical properties are strictly quantum mechanical and whose excitation spectrum is distinctively gapless. As a consequence, the charge current is largely uncorrelated from ionic currents and charge transport occurs without any significant mass displacement. The electronic spectrum of ionic conductors, instead, features an energy gap that constrains the electrons to remain in their instantaneous ground state at all times and their charge density and current to follow adiabatically the classical atomic motion. As a consequence, charge and mass currents are intrinsically entangled and charge transport cannot occur without mass convection.

Non-stoichiometric ionic conductors are somewhat intermediate between these two extrema. As the concentration of one the chemical species that make for the ionic components of the material is increased, it may happen that, not being compensated by ions of opposite charge, the chemical species in excess dissociates into an ionic moiety plus an unbound solvated electron. As the concentration of the excess species increases, the solvated electrons form an energy band that eventually merges into the unoccupied states of the stoichiometric system, thus turning the ionic conductor into a metal. Such a transition has been recently evinced by photo-electron spectroscopic measurements on alkali-metal solutions in liquid ammonia~\cite{buttersack2020photoelectron}. Before this critical concentration is reached, the excess electrons may coalesce into localized pairs that diffuse through the ionic matrix largely uncorrelated from the atomic motion, thus determining a transport regime where most of the charge is transported without appreciable mass displacement, while the system remains non-metallic.
This transport regime is often described in terms of the Marcus-Hush theory~\cite{marcus1985} of adiabatic electron-transfer, as resulting from an avoided crossing between potential energy surfaces. In this paper we have identified electron-transfer processes as a dynamical diagnostics of a topological anomaly in ACS, a manifestation of what we have called a breakdown of SA. While this anomaly is a blueprint of mass-less electron transfer, it does not provide any circumstantial description of the specific mechanisms by which the adiabatic transfer reaction takes place in realistic physical conditions.  On the other hand, SA breaking is a necessary condition for adiabatic electron-transfer processes to occur, thus indicating that the appearance of the latter is a dynamical manifestation of the former.

Of considerable interest is likewise the extension of our analysis to non-avoided crossings in the proximity of diabolical points in ACS, where nonadiabatic transitions are facilitated~\cite{yarkony1996diabolical} and geometric phases are known to play an important role in the dynamics~\cite{berry1984quantal, berry1984diabolical}.
The impact of non-trivial charge transport on heat transport is also of great significance. In particular, non-trivial charge transport may determine a high electric conductivity in electrolytes, not accompanied by a large heat conductivity, which is of potential interest for thermoelectric applications.

\begin{acknowledgments}
This work was partially funded by the EU through the \textsc{MaX} Centre of Excellence for supercomputing applications (Project No. 824143). We thank Luca Grisanti for insightful suggestions and a critical reading of the manuscript. FG thanks Stefano de Gironcoli and Luigi Grasselli for fruitful discussions.
\end{acknowledgments}

\appendix

\section{Computational details} \label{App:computational}

Electronic structure calculations and AIMD simulations are carried out within DFT using the plane-wave pseudopotential method with the \texttt{pw.x} and \texttt{cp.x} codes of the \qe package~\cite{quantum-espresso-1, quantum-espresso-2}. The transformation to the Wannier representation is performed, when needed, using the \wannier code~\cite{wannier90-1, wannier90-2}.

For the $\mathrm{H_3^+}$ system, the PBE$0$ hybrid functional~\cite{pbe0} is used in order to minimize self-interaction artifacts. A norm-conserving pseudopotential for H atoms has been generated to be consistent with the hybrid functional. The plane-wave kinetic-energy cutoff is set to $80\un{Ry}$ for wavefunctions and to $320\un{Ry}$ for both the charge density and non-local exchange operator.

For $\mathrm{K_3Cl}$, DFT calculations are performed at the GGA level in the PBE flavor~\cite{Perdew1996}. Norm-conserving pseudopotentials from the SG15 data set~\cite{ONCVpseudo,Schlipf2015} are used for K and Cl. The kinetic energy cutoff is set to $55$ and $220\un{Ry}$ for wave-functions and charge densities, respectively. Brillouin-zone (BZ) sampling is performed using a $6\!\times\!6\!\times\!6$ Monkhorst-Pack set of $k$-points~\cite{Pack1977}.

The computational parameters for $\mathrm{K_{33}Cl_{31}}$ are the same as for $\mathrm{K_3Cl}$ but for the BZ sampling, which is restricted to the $\Gamma$ point. The dynamics is carried out according to the Car-Parrinello Lagrangian scheme~\cite{Car1985} using a fictitious electronic mass ${\mu=400\,m_e}$, $m_e$ being the electronic physical mass, and a time-step of $0.36\un{fs}$; the current is sampled every 20 time steps.

\begin{figure}[t]
    \centering
    \includegraphics[width=0.9\columnwidth]{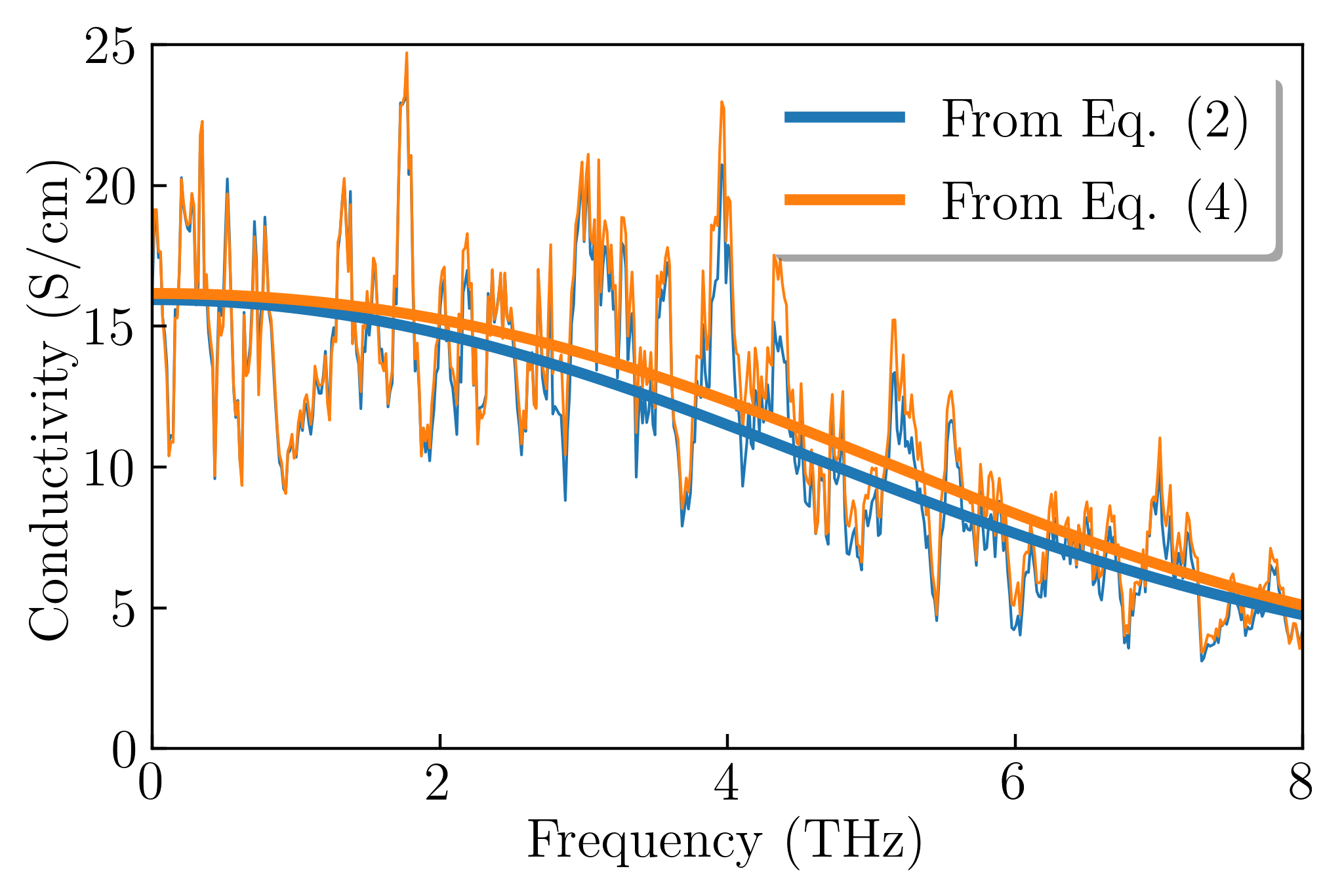}
    \caption{Low-frequency portion of the PSD of the displaced dipole computed according to Eq.~\eqref{eq:core + tutti WC} (blue) and Eq.~\eqref{eq:no + lb WC} (orange). The noisy lines are the window-filtered PSDs (with a window of 0.1 THz), while the smooth lines are the cepstral estimates.}
    \label{fig: cepstral}
\end{figure}

\section{Cepstral analysis} \label{App:cepstral}

The numerical values of the conductivities reported in the text are obtained from the \emph{cepstral} analysis of the electric current time-series~\cite{ercole2017accurate,bertossa}, which allows one to obtain accurate estimates of both transport coefficients and their statistical accuracy. In a nutshell, cepstral analysis is based on the Wiener-Kintchine theorem~\cite{champeney1989handbook}, which allows one to express the conductivity as
\begin{equation}
    \sigma = \frac{1}{2 \kb T L^3} S(\omega=0), \label{eq:WK}
\end{equation}
where $S(\omega)$ is the power spectral density (PSD) of the electric current. The electric current, $ J(t)=\frac{d{\mu}(t)}{dt}$, where $\mu$ is any Cartesian component of the displaced dipole defined in Eqs.~\eqref{eq:core + tutti WC} or~\eqref{eq:no + lb WC}, is an extensive quantity whose density has correlations that are usually short-ranged. Therefore, according to the central limit theorem, $J(t)$ is a Gaussian process whose Fourier transform, ${\tilde{J}(\omega)}$, is normally distributed and such that $\tilde{J}(\omega)$ is uncorrelated from ${\tilde{J}(\omega')}$ for ${\omega \neq \omega'}$ in the large-time limit.

For any discrete realization of the continuous current process, $\hat{J}_n = J(n\varepsilon)$, with ${n=1,\ldots,N}$, we define its discrete Fourier transform, $\hat{\tilde{J}}_k=\sum_n J_n \mathrm e^{i2\pi\frac{ kn}{N}}$ and the periodogram as the random variate
\begin{equation}
    \hat{S}_k=\frac{\varepsilon}{N}\abs{\hat{\tilde{J}}_k}^2.
\end{equation}
Due to the Gaussian nature of $\hat{J}_n$, $\hat{S}_k$ is proportional to a $\chi^2$ random variate, and its expected value is the PSD in Eq.~\eqref{eq:WK}. The application of a low-pass filter to the logarithm of this quantity, $\hat{L}_k$, yields a consistent estimator of the logarithm of the conductivity; the low-pass filter consists in the retention of a number $P^\star \ll N/2$ of (inverse) Fourier components of $\hat{L}_k$, whose value is chosen according to the Akaike Information Criterion (AIC)~\cite{akaike} of model selection. In order to limit the analysis to an appropriate low-frequency portion of the entire PSD, it is expedient to re-sample the electric current time-series with a rate corresponding to an effective Nyqvist frequency, $f^\star$. In Fig.~\ref{fig: cepstral} we show the filtered PSDs for the electric currents computed both with Eq.~\eqref{eq:core + tutti WC} and~\eqref{eq:no + lb WC}. The cepstral analysis was performed with the \texttt{Thermocepstrum} code~\cite{thermocepstrum,Baroni2018}. For both currents, the value chosen by the AIC is $P^\star=8$. The estimated conductivities depend very little on the re-sampling frequency, which has been set to $f^\star=10\un{THz}$.


%

\end{document}